\documentclass[12pt]{cernart}

\usepackage{graphicx}
\usepackage{amsmath}
\usepackage{amssymb}
\usepackage{times}
\usepackage{cite}
\usepackage{multirow}
\usepackage[table]{xcolor}
\usepackage{appendix}

\begin{document}
\begin{titlepage}
%\vspace{3.1cm}
\flushright{
\includegraphics[width=2cm]{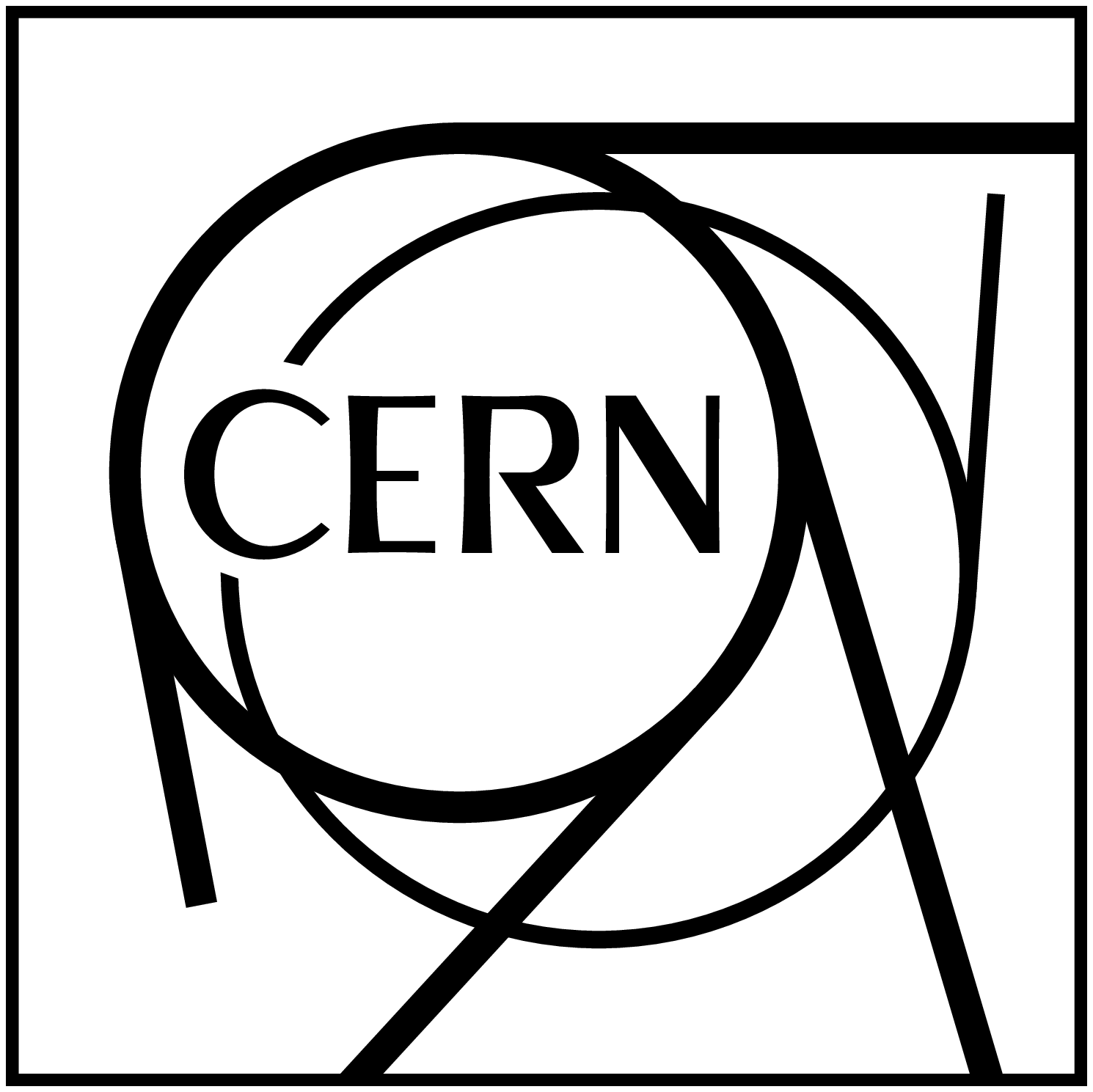}\\
CERN-PH-EP-2013-225 \\
06 January 2014 \\
}
%\docnum{CERN-PH-EP-2013-225}
%\date{06 January 2014} 
%\docnum{draft 0.9}
%\date{}
%
\vspace{2mm}
\title{\large{Comments concerning the paper "Measurement of negatively charged
              pion spectra in inelastic p+p interactions at 20, 31, 40, 80 and 158~GeV/c" 
              by the NA61 collaboration}}
\begin{Authlist}
\vspace{2mm}
\noindent
 H.~G.~Fischer\Iref{cern}, 
 M.~Makariev\Iref{inrne},
 D.~Varga\Iref{kfki},
 S.~Wenig\Iref{cern}
\end{Authlist}

%{\it (The NA49 Collaboration)}  \\
\vspace*{5mm}
\Institute{kfki}{KFKI Research Institute for Particle and Nuclear Physics, Budapest, Hungary}
\Institute{cern}{CERN, Geneva, Switzerland}
\Institute{inrne}{Institute for Nuclear Research and Nuclear Energy, BAS, Sofia, Bulgaria}
%
%%\vspace{10mm}%\vspace*{5cm}
%\begin{center}
%{\small{\it to be published in EPJC }}
%\end{center}
\vspace*{10mm} 
%\clearpage

\begin{abstract}
\vspace{-3mm}
New data from the NA61 collaboration on the production of negative 
pions in p+p interactions at beam momenta between 20 and 158~GeV/c
are critically compared to available results in the same energy 
range. It is concluded that the NA61 data show some discrepancies with the
previous results. This concerns in particular the total
yields, the $p_T$ integrated rapidity distributions and the double
differential cross sections.
\end{abstract}
%\vspace*{5mm}
 
\clearpage
\end{titlepage}

%
% ****************************** Section 1 ****************************
%
\section{Introduction} 
\vspace{3mm}
\label{sec:intro}

The study of inclusive hadron production in p+p interactions has
a long history. It has closely followed the steady evolution of 
particle accelerators towards higher interaction energies as well as
the development of novel detector technologies. Although the
measurement of particle yields in well-defined regions of the
available phase space might seem to be a rather trivial undertaking, 
the achievement of high precision presents a real challenge to
the experimentalist in terms of absolute normalization and the
control of systematic effects, in particular concerning the
necessary corrections for detector and accelerator connected
effects. The term of "precision" has to be defined here very
carefully. In fact a survey of the existing data reveals that
it is the evaluation of systematic errors rather than the available
event statistics that is the limiting factor. Indeed only a few
experiments have achieved cross section measurements with a
systematic uncertainty in the one to two percent range. It is
on this level that one may talk about "high precision" data,
and this performance has shown no real evolution over the past few
decades of experimental work. It is by no means true that the
use of state-of-the-art detector technology would automatically
ensure superior data quality. Indeed any new attempt at producing
"precision" data has to take full account of the preceding work.
It is in this sense that this paper presents a critical review
of new data on $\pi^-$ yields produced by the NA61 collaboration
in the range of beam momenta from 20 to 158~GeV/c \cite{na61}. In the
introduction to their publication the authors claim that "the
available data concern mainly basic features of unidentified
charged hadrons and they are sparse. Many needed results on
hadron spectra are missing". And they continue: "Thus new high
precision measurements on hadron production properties in p+p
interactions are necessary...". These claims are put to a stringent
test in the present comments, and it will be shown that the new
data are by no means superior to existing results. In fact the
NA61 results are found to be less precise than a set of reference experiments
over the full range of the studied beam momenta.

This paper is arranged as follows. After an introduction to the
experimental situation concerning inclusive pion production, 
a set of reference experiments in the range of beam momenta 
from 12 to 158~GeV/c is recalled in Sect.~\ref{sec:ref_exp}. 
Section~\ref{sec:na61} describes some aspects of the new NA61 results. A detailed
comparison of these data with the published reference
experiments follows for the total $\pi^-$ yields in
Sect.~\ref{sec:tot_pim}, for the $p_T$ integrated rapidity distributions in
Sect.~\ref{sec:rap_dist} and for the double differential cross sections
in Sect.~\ref{sec:differential}. Some remarks concerning normalization problems
in Sect.~\ref{sec:corr} conclude the paper.  

%
% ****************************** Section 2 ****************************
%
\section{The experimental situation} 
\vspace{3mm}
\label{sec:exp_sit}

Most early experiments studying secondary hadron production in
p+p interactions have been using hydrogen bubble chambers. This
technique has a decisive advantage over all other detectors in
terms of the continuous phase space coverage, perfect absolute 
normalization and only small corrections for detector related 
effects. Hence it offers superior performance concerning the 
systematic uncertainties.

For beam momenta up to about 5~GeV/c, the study of fully constrained
exclusive final states has been possible using kinematical fitting. 
Above this limit, the transition to inclusive single particle cross
sections had to be accepted due to the increased
total multiplicity including numerous non-detected neutral hadrons.
This accentuated the inherent weakness of bubble chamber data in
terms of particle identification and of the rather sharp limitation
in the size of the obtainable data samples. For the study of $\pi^-$
production however the former limitation is of less importance 
in energy regions where the K$^-$/$\pi^-$ and $\overline{\textrm{p}}$/$\pi^-$ ratios remain limited
to the few percent level and methods of precise corrections have
been developed, especially making use of the complete phase space
coverage of the bubble chamber results. These methods have been
employed up to beam momenta of about 70~GeV/c.

The alternative detector technique using small solid angle 
magnetic spectrometers allows for complete particle identification
using for instance Cerenkov or time-of-flight detectors. Its
weaknesses lie in the increased problems around calibration,
normalization and corrections as well as in the generally very 
limited and fractional coverage of the production phase space. 
In fact not a single spectrometer experiment has been covering the
complete phase space in one set-up. The introduction
of the Time Projection Chamber (TPC) as a large solid angle detector
permitting combined precision tracking and particle identification
has remedied at least part of these problems. It offers, in fixed 
target experiments, identification over a major fraction of phase 
space via energy loss measurement with the exception of the 
cross-over region between the energy loss of the different particle 
species at about 1--3~GeV/c laboratory momentum. This corresponds 
to different areas in accessible phase space as a function of beam 
momentum. In addition TPC detectors need, as they are used in 
conjunction with targets of beam interaction lengths of at most 
a few percent, external triggering which introduces non-trivial and sizeable corrections. 

In the following sections a number of existing results in the
range of beam momenta covered by the NA61 experiment will be 
used as a reference for a detailed comparison with the new data.
In this comparison the following definitions and variables will
be used:

\begin{itemize}
  \setlength{\itemsep}{0.5mm}
  \item laboratory beam momentum $p_{\textrm{beam}}$ (GeV/c)  
  \item transverse momentum $p_T$ (GeV/c)
  \item cms energy squared $s = 2m_p + 2m_pp_{\textrm{beam}}$ ($m_p$ = proton mass) (GeV$^2$)
  \item cms rapidity $y = 0.5 \ln(( E + p_L )/( E - p_L ))$ ($p_L$ cms longitudinal momentum)
  \item $\langle n_{\pi^-} \rangle$ average number of $\pi^-$ per inelastic event
  \item transverse mass $m_T = \sqrt{p_T^2 + m_\pi^2}$ ($m_\pi$ = pion mass)
  \item invariant inclusive cross section $f(y,p_T) = \frac{1}{\pi} \frac{d^2\sigma}{dydp_T^2}$
  \item pion density per inelastic event $d^2n/dydp_T$
  \item $p_T$ integrated pion density $dn/dy$ per inelastic event
\end{itemize}

%
% ****************************** Section 3 ****************************
%
\section{The reference experiments} 
\vspace{3mm}
\label{sec:ref_exp}

%
% ****************************** Section 3.1 ****************************
%
\subsection{Beam momenta, event samples and systematic uncertainties} 
\vspace{3mm}
\label{sec:beam}

A series of eight published experiments have been selected in the 
range 12~$< p_{\textrm{beam}} <$~158~GeV/c. These are completely bracketing the NA61
beam momenta. Some information concerning the beam momenta, cms
energies and the size of the event samples are summarized in Table~\ref{tab:beam}.
  
% Table 1
\begin{table}[h]
%\renewcommand{\tabcolsep}{0.14pc} 
%\renewcommand{\arraystretch}{1.0}
%\footnotesize
%\scriptsize
 \begin{center}
  \begin{tabular}{|ccccccccc|}
  \hline
  $p_{\textrm{beam}}$ (GeV/c) &    12  &  19  &  24  &   28.5  &  32  &  69  &  100  &  158  \\
  reference                   &    \cite{blobel}  &  \cite{boggild}  & \cite{blobel}  &  \cite{sims}  
                              &  \cite{zabrodin}  &  \cite{ammosov}  & \cite{morse}  & \cite{pp_pion}  \\
  $\sqrt{s}$(GeV)  &   4.92 &   6.12  &  6.84  &  7.43  &  7.85 &  11.45 &  13.76 &  17.27  \\
  events used   &   175k   &   8k  &  100k   &  30k  &  100k   &   8k   &   4k  &  4700k  \\
  detector      &    HBC   &  HBC  &   HBC  &   HBC  &   HBC  &   HBC  &   HBC   &   TPC   \\
  \hline
  \end{tabular}
  \end{center}
  \caption{Selection of reference experiments for $\pi^-$ production
           in p+p interactions, giving beam momentum, $\sqrt{s}$, the
           number of events available and the type of detector (HBC =
           hydrogen bubble chamber)}
  \label{tab:beam}
\end{table}

It should be mentioned here that the event numbers of the bubble
chamber experiments, especially refs. \cite{blobel,zabrodin} are rather sizeable 
for this detector technique and well adapted to the systematic
errors, see below.

All reference experiments have published $\langle \pi^- \rangle$ and rapidity
distributions $dn/dy$. In addition double differential cross
sections $f(y,p_T)$ are available in \cite{blobel,zabrodin,ammosov,pp_pion}.
The bubble chamber experiments \cite{blobel,boggild,sims,zabrodin,ammosov,morse} achieve measurements of the 
total charged multiplicity $\langle n_{\textrm{ch}} \rangle$ with a precision of about 1\%.
For the event numbers given in Table~\ref{tab:beam} this uncertainty is in
most cases not governed by the statistical errors. For $\langle n_{\pi^-} \rangle$ 
the given errors are about 2\%, again not dominated by statistics. 
The TPC experiment NA49 \cite{pp_pion} gives an overall systematic uncertainty 
of 2\%. This means that all reference data offer a systematic 
precision on the 2\% level.

%
% ****************************** Section 3.2 ****************************
%
\subsection{Particle identification} 
\vspace{3mm}
\label{sec:pid}

\begin{figure}[h!]
  \begin{center}
  	\includegraphics[width=6.cm]{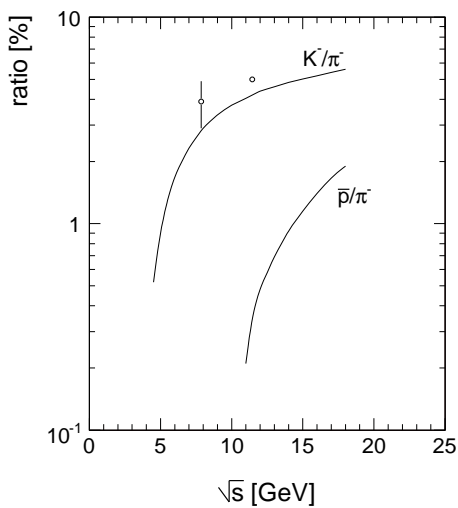}
      \caption{K$^-$/$\pi^-$ and $\overline{\textrm{p}}$/$pi^-$ ratios as a function of
               $\sqrt{s}$. The full lines correspond to the new determination
               of the energy dependence of K$^-$ \cite{pp_kaon} and $\overline{\textrm{p}}$ \cite{pp_proton} yields
               by the NA49 collaboration. The measurement of the K$^-$/$\pi^-$
               ratio at 32~GeV/c beam momentum and the correction deduced
               from the measured K$^0_S$ yields at 69~GeV/c are given as data points}
  	  \label{fig:ratio_tot}
  \end{center}
\end{figure}

The situation concerning the contribution of K$^-$ and anti-protons
is clarified in Fig.~\ref{fig:ratio_tot}.

%        Fig.1 

It is apparent from Fig.~\ref{fig:ratio_tot} that in relation to an overall systematic
uncertainty of 2\% the K$^-$ and $\overline{\textrm{p}}$ contributions are of importance
above $\sqrt{s} \gtrsim$~5~GeV and 14~GeV, respectively. The K$^-$ and $\overline{\textrm{p}}$ 
yields are directly determined by energy loss analysis in the
NA49 experiment \cite{pp_kaon,pp_proton}. The bubble chamber experiments at 32 and
and 69~GeV/c beam momentum have used either a direct determination 
of the K$^-$ yields \cite{zabrodin} or corrections derived from the measurement 
of K$^0_S$ cross sections \cite{ammosov} in order to take account of the kaon 
contributions. Reference \cite{zabrodin} measures a K$^-$/$\pi^-$ ratio of 4\%$\pm$1.5\%, ref. \cite{ammosov} 
gives a correction of 5--6\%, see Fig.~\ref{fig:ratio_tot}. It may therefore be concluded 
that at least to an accuracy within the given systematic errors the 
identification of $\pi^-$ yields has been achieved. For the group of 
measurements at beam momenta between 19 and 28.5~GeV/c \cite{blobel,boggild,sims} no 
subtraction of K$^-$ contributions has been attempted in the published 
results. As in this energy region the K$^-$/$\pi^-$ ratio is of order 
1.7--2.2\%, a corresponding reduction of the published yields has been 
performed in the subsequent sections of this paper. The data \cite{morse}
at $p_{\textrm{beam}}$~=~100~GeV/c need a somewhat bigger downward correction of 
about 5\%, see Fig.~\ref{fig:ratio_tot}. Here the internal consistency with the other
reference data may be controlled regarding the energy dependence of
the total yields and of the rapidity distributions, Sects.~\ref{sec:tot_pim} and \ref{sec:rap_dist}.

%
% ****************************** Section 4 ****************************
%
\section{The NA61 experiment} 
\vspace{3mm}
\label{sec:na61}

%
% ****************************** Section 4.1 ****************************
%
\subsection{Beam momenta, event samples and systematic uncertainties} 
\vspace{3mm}
\label{sec:na61_beam}

Some information about the NA61 data \cite{na61} concerning the beam momenta
and the number of used events is given in Table~\ref{tab:na61_beam}.

% Table 2
\begin{table}[h]
\renewcommand{\arraystretch}{1.2}
 \begin{center}
  \begin{tabular}{|cccccc|}
  \hline
   $p_{\textrm{beam}}$ (GeV/c)  &   20   &   31   &   40   &   80    &   158    \\ 
   $\sqrt{s}$(GeV)              &  6.27  &  7.74  &  8.76  &  12.32  &  17.27   \\
   events used                  &  233k  &  843k  & 1580k  &  1540k  &  1650k   \\
  \hline
  \end{tabular}
  \end{center}
  \caption{Beam momenta, $\sqrt{s}$ and the available event samples of the NA61 experiment}
  \label{tab:na61_beam}
\end{table}

\begin{figure}[b!]
  \begin{center}
  	\includegraphics[width=15cm]{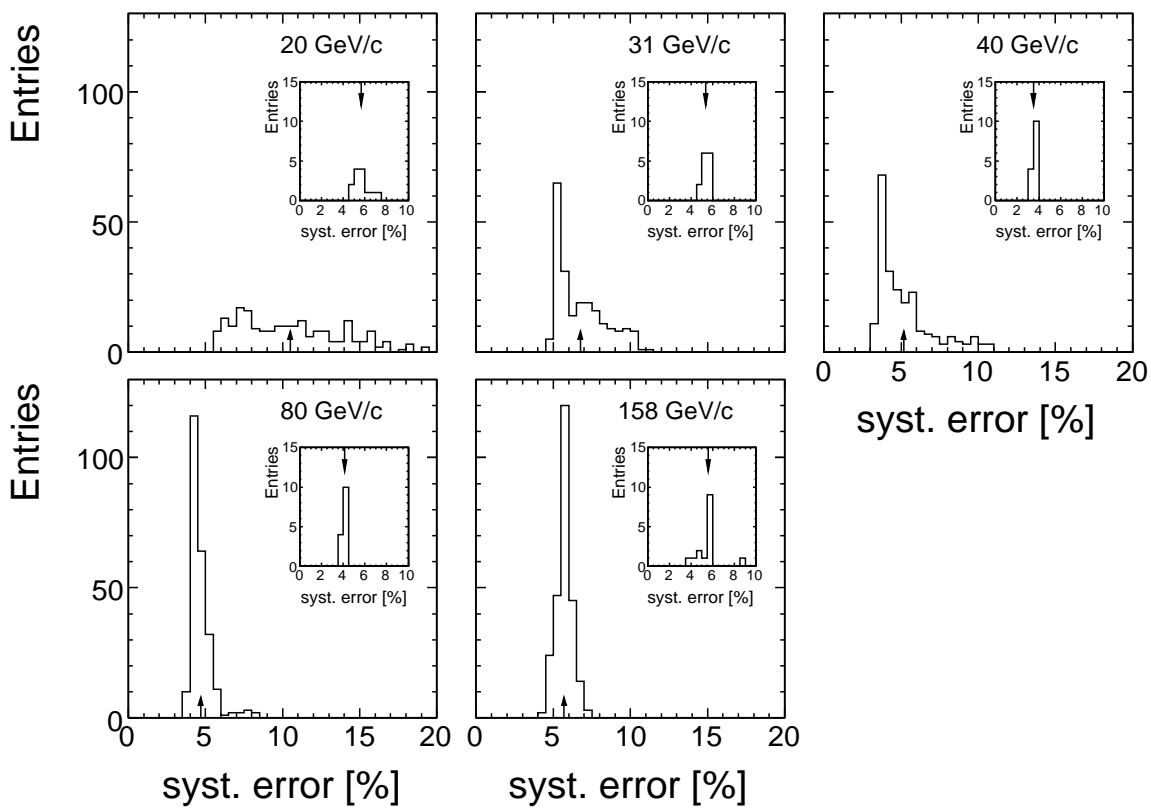}
      \caption{Bin-by-bin distributions of the systematic errors
               of the NA61 experiment for the double differential yields $d^2n/dydp_T$ at 
               the five beam momenta between 20 and 158~GeV/c. In the inserts the
               systematic errors of the respective $p_T$ integrated distributions $dn/dy$ are presented. 
               The corresponding mean values are indicated by arrows}
  	  \label{fig:syst_dist}
  \end{center}
\end{figure}

The available event numbers are superior to the ones of the
bubble chamber experiments \cite{blobel,boggild,sims,zabrodin,ammosov,morse}, however this advantage is
offset by the very sizeable systematic uncertainties.
The bin-by-bin distributions of the given systematic errors are
shown in Fig.~\ref{fig:syst_dist}.

%        Fig.2 

Evidently the NA61 data feature rather important systematic
uncertainties on the bin-to-bin level, with a minimum at about 5--6\% and maximum values
reaching up to 20\% at low beam momenta. This means that these
errors are by factors of 3--10 above the ones published by the
reference experiments. The strong difference between the uncertainties
of the $p_T$ integrated quantities, see inserts in Fig.~\ref{fig:syst_dist},
and the ones given for the individual bins are somewhat surprising. They would
indicate very strong bin-to-bin correlations due to finite transverse momentum
resolution. Given the event statistics of order
1--2 Mevents it may be stated that the experimental errors are
governed by systematics rather than by statistical fluctuations.

%
% ****************************** Section 4.2 ****************************
%
\subsection{Particle identification} 
\vspace{3mm}
\label{sec:na61_pid}

NA61 does not attempt to use the superior particle identification 
capabilities offered by its TPC detectors \cite{pp_pion,pp_kaon,pp_proton} with the exception
of a correction for electron contamination. Instead, the yields
of all negative hadrons (h$^-$) are determined and a correction for
K$^-$ and $\overline{\textrm{p}}$ contributions is applied using a microscopic production
model \cite{epos}. No quantitative information is given in \cite{na61} as to the
precision to which this model might describe the K$^-$ and $\overline{\textrm{p}}$ yields
as functions of $p_{\textrm{beam}}$ and rapidity. A detailed comparison would
be particularly mandatory in view of the recent work on the 
$s$-dependence of kaon and baryon production published by NA49 \cite{pp_kaon,pp_proton}.

%
% ****************************** Section 5 ****************************
%
\section{Total $\pi^-$ yields} 
\vspace{3mm}
\label{sec:tot_pim}

For a precise comparison of $\langle \pi^- \rangle$ the following reference data
from experiments only giving the total $\pi^-$ yields, Table~\ref{tab:ref_beam}, 
have been added to the list of Table~\ref{tab:beam}.

% Table 3
\begin{table}[h]
\footnotesize
\renewcommand{\arraystretch}{1.4}
 \begin{center}
  \begin{tabular}{|cccccccccccc|}
  \hline
   $p_{\textrm{beam}}$ (GeV/c)  &  2.23 & 2.78 & 3.68 &  5.50  &  6.60  &  60  &  100  &  102  &  205  &  300   &  400  \\   
    ref.                        &  \cite{eisner} &  \cite{pickup} & \cite{xx} & \cite{alexander} & \cite{gellert}  
                                & \cite{bromberg} & \cite{erwin} & \cite{bromberg1} & \cite{barish} & \cite{firestone} 
                                & \cite{bromberg1,kass,ehs}  \\
  \hline
  \end{tabular}
  \end{center}
  \caption{Beam momenta for the additional reference data for $\langle \pi^- \rangle$,
           refs. \cite{eisner,pickup,xx,alexander,gellert,bromberg,erwin,bromberg1,barish,firestone,kass,ehs}}
  \label{tab:ref_beam}
\end{table}

\begin{figure}[h!]
  \begin{center}
  	\includegraphics[width=15.cm]{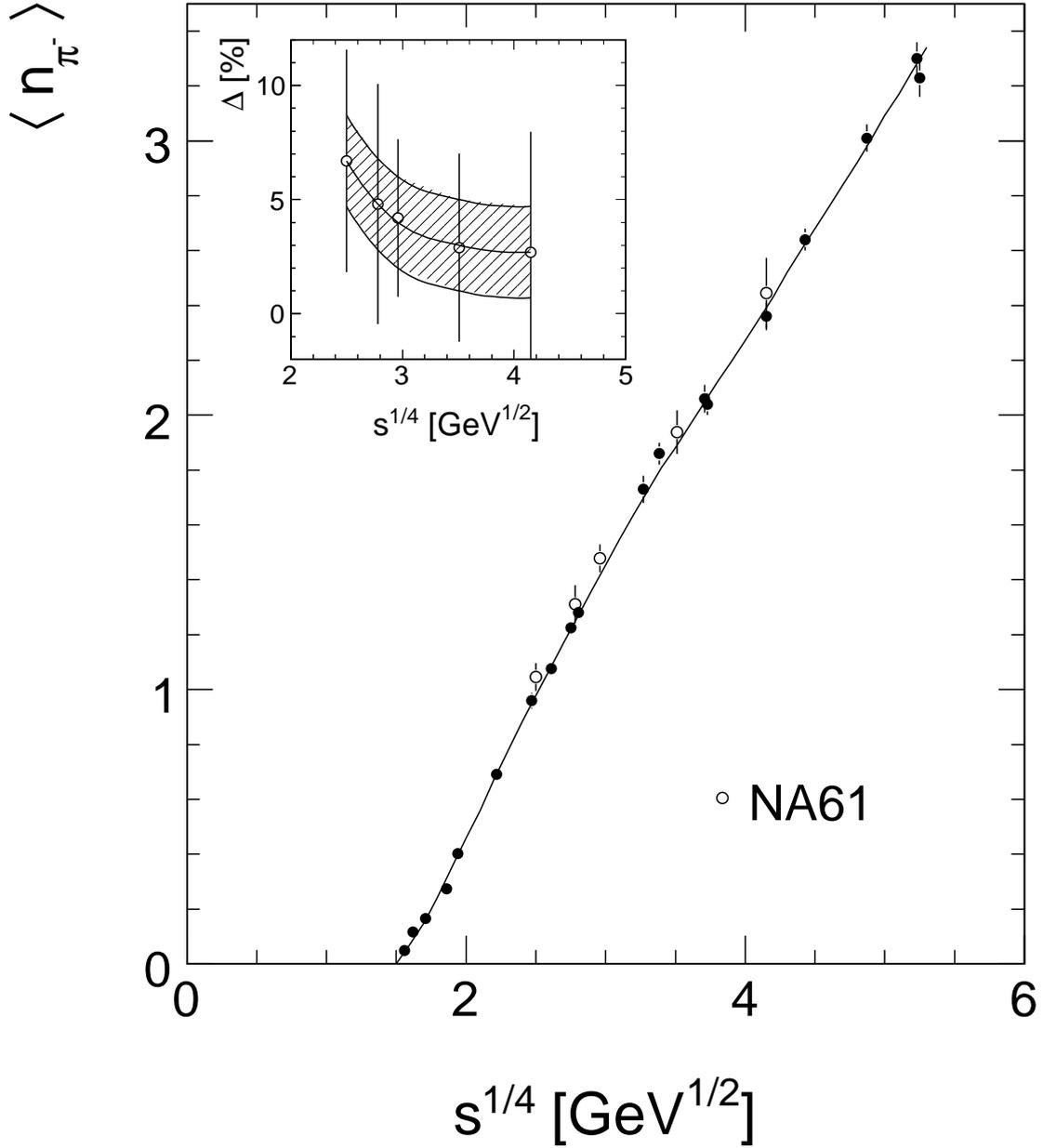}
      \caption{Average $\pi^-$ multiplicities as a function of
               s$^{1/4}$. The reference experiments are given as
               closed circles. The full line corresponds to an
               eyeball fit through the reference data. Open
               circles: NA61 results with their published systematic
               errors. The insert gives the percent deviations 
               of the NA61 data from the interpolation of the
               reference data. The line is drawn to guide the eye. The error bars are
               quoted by NA61 \cite{na61}, the shaded area corresponds to the systematic
               uncertainties of the reference experiments}
  	  \label{fig:meanpi}
  \end{center}
\end{figure}

With the exception of the low-energy measurements \cite{eisner,pickup,xx,alexander,gellert} and the
EHS experiment \cite{ehs} the data are obtained from $\langle h^- \rangle$. They have been 
corrected for K$^-$ and $\overline{\textrm{p}}$ contributions using the particle ratios 
of Fig.~\ref{fig:ratio_tot}. The resulting average $\pi^-$ yields are given in Fig.~\ref{fig:meanpi} as 
a function of s$^{1/4}$.

%        Fig.3 
  
It is evident from Fig.~\ref{fig:meanpi} that the NA61 data lie above the reference
results for all values of $p_{\textrm{beam}}$. The percent deviation increases
with decreasing beam momentum and reaches 7\% at $p_{\textrm{beam}}$~=~20~GeV/c
(insert of Fig.~\ref{fig:meanpi}).
 
The deviations which are visible in Fig.~\ref{fig:meanpi} have direct
consequences for the overall precision of the NA61 experiment. If
the $\pi^-$ yield is increased, the h$^+$ yield has to follow upwards,
albeit with a smaller percentage, in order to not violate charge 
conservation. This will in turn offset the total charged multiplicity
which is measured to a one percent systematic accuracy by most
reference experiments. 
 
%
% ****************************** Section 6 ****************************
%
\section{Rapidity distributions $dn/dy$} 
\vspace{3mm}
\label{sec:rap_dist}

%
% ****************************** Section 6.1 ****************************
%
\subsection{Reference rapidity density distributions as a function of $y$ and $p_{\textrm{beam}}$} 
\vspace{3mm}
\label{sec:ref_rap_dist}

\begin{figure}[h!]
  \begin{center}
  	\includegraphics[width=15cm]{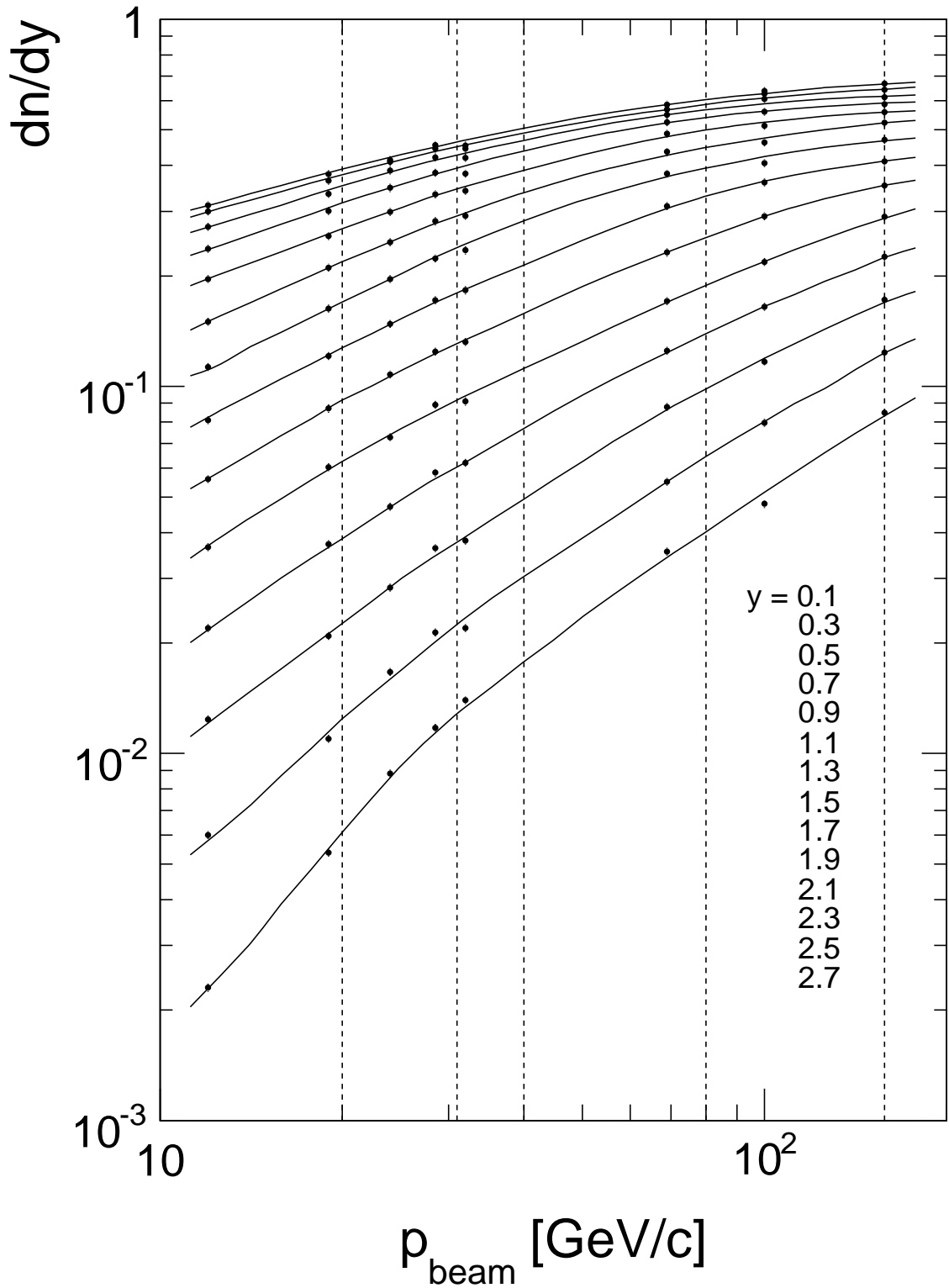}
      \caption{Rapidity densities $dn/dy$ from eight reference
               experiments (full dots) as a function of $p_{\textrm{beam}}$ for
               rapidities between 0.1 and 2.7. The full lines constitute an 
               eyeball interpolation. The positions of the five beam momenta
               measured by NA61 are presented as the vertical broken lines}
  	  \label{fig:dndy_bc}
  \end{center}
\end{figure}

The single differential, $p_T$ integrated rapidity distributions 
$dn/dy(y,p_{\textrm{beam}})$ constitute a next step towards a more detailed inspection
of the NA61 results. All reference experiments cited in Table~\ref{tab:beam}
have published rapidity distributions. These have been interpolated
to the rapidity values published in \cite{na61}. The resulting rapidity
densities are shown in Fig.~\ref{fig:dndy_bc} as a function of beam momentum for
the rapidity range from 0.1 to 2.7. The data points (full dots)
are interpolated in $p_{\textrm{beam}}$ by the full lines.

%        Fig.4 

\begin{figure}[h!]
  \begin{center}
  	\includegraphics[width=8.cm]{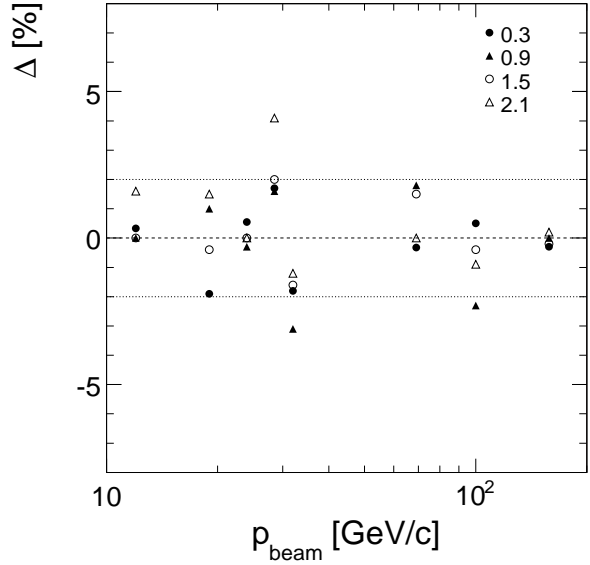}
      \caption{Point-by point deviation of the reference data
               from the interpolated $p_{\textrm{beam}}$ dependence, given in percent,
               as a function of $p_{\textrm{beam}}$ for fixed rapidities between 0.3 and 2.1}
  	  \label{fig:dndy_diff}
  \end{center}
\end{figure}

Evidently the eight cited experiments constitute a data sample
which is internally consistent and compatible with a continuous
and smooth dependence both on $p_{\textrm{beam}}$ and on rapidity. The 
deviations of the data points from the interpolation are consistent
with the systematic uncertainty of 2\% given by all experiments.
This is exemplified in Fig.~\ref{fig:dndy_diff} where the percent deviations of
the data points with respect to the interpolated $p_{\textrm{beam}}$ dependence
are shown for four rapidity values between 0.3 and 2.1.

%        Fig.5 

%
% ****************************** Section 6.2 ****************************
%
\subsection{Comparison to the NA61 results} 
\vspace{3mm}
\label{sec:comp_rap_dist}

\begin{figure}[h!]
  \begin{center}
  	\includegraphics[width=11.5cm]{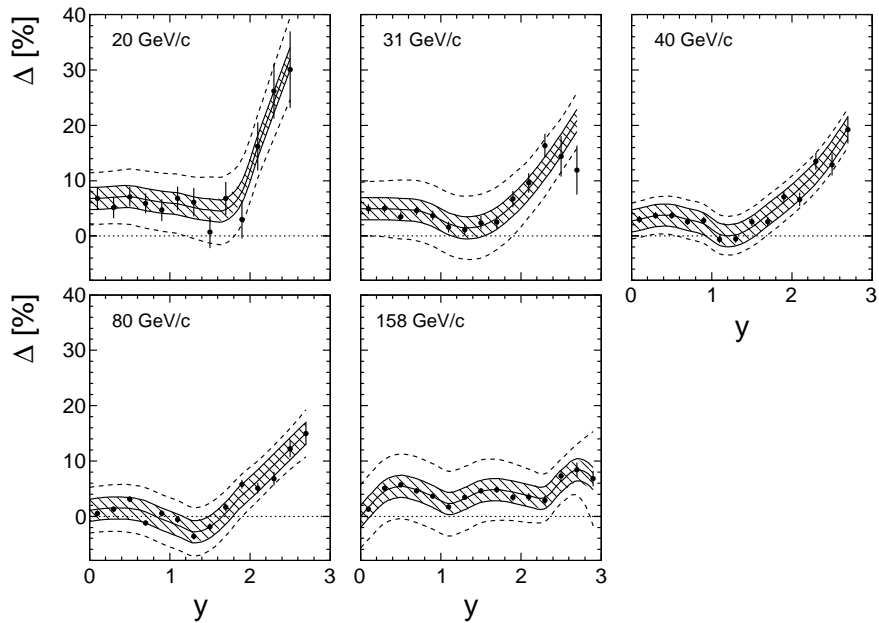}
      \caption{Relative difference between the NA61 results and
               the interpolated reference data in percent as a function of rapidity,
               for the five values of $p_{\textrm{beam}}$ given by NA61.
               The lines through the points are drawn to guide the eye. The shaded 
               areas and the dashed lines represent the systematic errors of the reference
               data and NA61 data, respectively}
  	  \label{fig:dndy_na61diff}
  \end{center}
\end{figure}

The global data sample contained in Fig.~\ref{fig:dndy_bc} may now be compared
to the NA61 results at the five respective beam momenta between
20 and 158~GeV/c. This comparison is presented in the five
panels of Fig.~\ref{fig:dndy_na61diff} giving the relative difference in percent between
the NA61 results and the reference data interpolation to the
NA61 beam momenta.

%        Fig.6 

Very substantial deviations appear in this comparison of the
single differential distributions. There is in general a structure
with positive deviations at low and high rapidity and an
intermediate minimum around $y$~=~1.3--1.5. The minimum becomes
slightly negative at $p_{\textrm{beam}}$~=~40 and 80~GeV/c. The positive
deviations reach typical values of about 6\% at low $y$ and between
10 and 30\% at $y >$~2 with a clear dependence on beam momentum.
This corresponds to discrepancies of more than 4 standard deviations
even compared to the larger NA61 errors.

%
% ****************************** Section 6.3 ****************************
%
\subsection{Gaussian fits} 
\vspace{3mm}
\label{sec:gaus_fits}

\begin{figure}[h!]
  \begin{center}
  	\includegraphics[width=10.5cm]{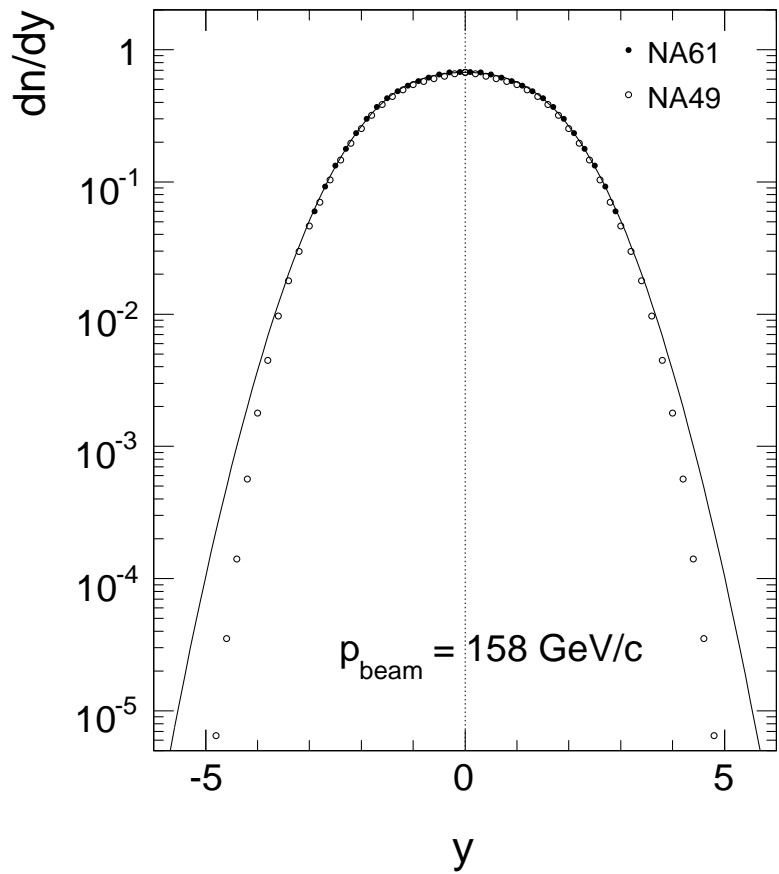}
      \caption{Rapidity density at $p_{\textrm{beam}}$~=~158~GeV/c as a
               function of rapidity. Full dots: NA61 data. Open circles:
               NA49 data. The full line represents the data parametrization
               with two symmetrically displaced Gaussians using the
               parameters given in \cite{na61}}
  	  \label{fig:ydist}
  \end{center}
\end{figure}

The fitting of rapidity distributions with two symmetrically
displaced Gaussians appears to be a standard procedure of NA61
in order to represent their data both for the elementary p+p
and for A+A interactions. It is of course not to be expected
that the physics of soft hadron production would satisfy such
a simplistic arithmetic parametrization. This is demonstrated
in Fig.~\ref{fig:ydist} which shows the comparison between the double 
Gaussian fit and the NA61 data (full dots) and the published
NA49 data (open circles) at 158~GeV/c beam momentum. 

%        Fig.7 

\begin{figure}[h!]
  \begin{center}
  	\includegraphics[width=12cm]{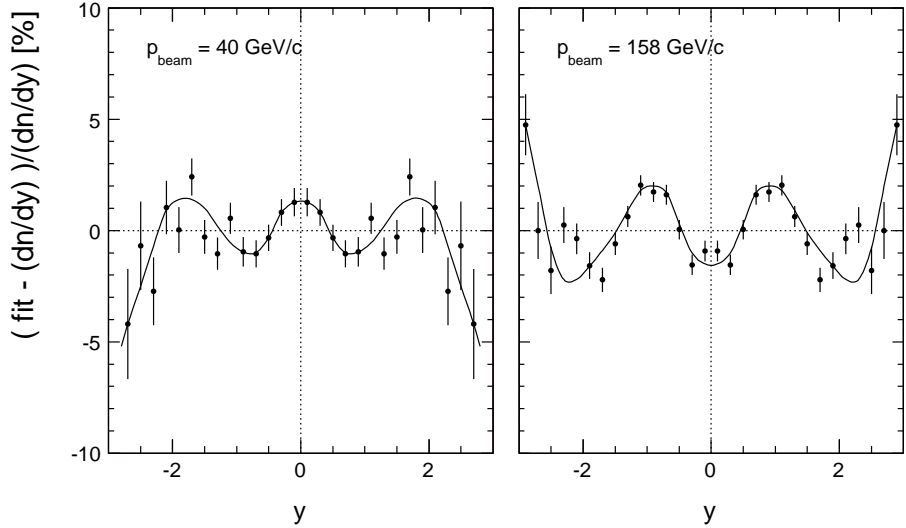}
      \caption{Deviations of the NA61 data from the Gaussian fit
               as a function of rapidity in percent at $p_{\textrm{beam}}$~=~40~GeV/c, 
               and 158~GeV/c. The lines are drawn to guide the eye}
  	  \label{fig:fit_diff}
  \end{center}
\end{figure}

Beyond the rapidity range covered by the NA61 data, the 
parametrization diverges from the published data by a factor
of 2 at $y$~=~4 and by more than one order of magnitude at 
$y$~=~4.8. But also within the rapidity range of the NA61
measurements there are important systematic differences between
data and fit as presented in Fig.~\ref{fig:fit_diff} for the beam momenta of
40 and 158~GeV/c.

%        Fig.8 

Evidently the fits introduce systematic distortions of up to $\pm$5\%
compared to the data. It has been repeatedly demonstrated in
the NA49 publications \cite{pp_pion,pp_kaon,pp_proton} that arithmetic formulations are
not advisable for data with a precision on the few percent level. 

%
% ****************************** Section 7 ****************************
%
\section{Double differential cross sections} 
\vspace{3mm}
\label{sec:differential}

A final step in the critical assessment of the NA61 results 
consists in scrutinizing the double differential $\pi^-$ densities
$d^2n/dydp_T$ in comparison to the reference experiments that have
given cross sections as a function of both y and $p_T$.

%
% ****************************** Section 7.1 ****************************
%
\subsection{Comparison to NA49 at $p_{\textrm{beam}}$~=~158~GeV/c} 
\vspace{3mm}
\label{sec:comp_na49}

\begin{figure}[h!]
  \begin{center}
  	\includegraphics[width=12.5cm]{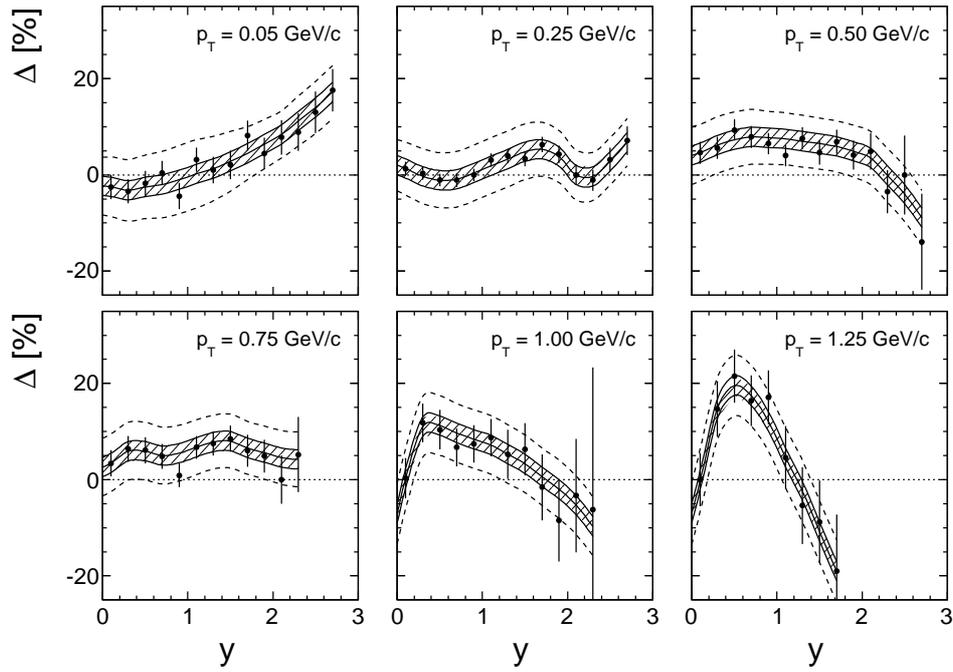}
      \caption{Relative differences between NA61 and NA49 results
               in percent of the NA49 cross sections for 6 values of $p_T$,
               as a function of rapidity. The error bars correspond to the
               statistical errors of NA61. The full lines are drawn to
               guide the eye. The shaded 
               areas and the dashed lines represent the systematic errors of the reference
               data and NA61 data, respectively}
  	  \label{fig:na61_na49_comp}
  \end{center}
\end{figure}

NA49 has published \cite{pp_pion} detailed interpolated invariant cross sections
$f(x_F,p_T)$ as functions of Feynman $x_F$ and $p_T$. These data are given
in steps of 0.05~GeV/c in $p_T$ and may be readily interpolated to the 
rapidity values chosen by NA61. The corresponding relative differences,
given in percent of the NA49 cross sections, are shown in Fig.~\ref{fig:na61_na49_comp} for
6 values of $p_T$ as a function of rapidity.

%        Fig.9 

Very substantial deviations of the NA61 data from the NA49 results
are visible in Fig.~\ref{fig:na61_na49_comp}. These deviations show a systematic
behaviour, reaching up to $\pm$20\% of the NA49 values.

%
% ****************************** Section 7.2 ****************************
%
\subsection{Comparison to the bubble chamber reference \cite{blobel}} 
\vspace{3mm}
\label{sec:comp_bc}

\begin{figure}[h!]
  \begin{center}
  	\includegraphics[width=12.5cm]{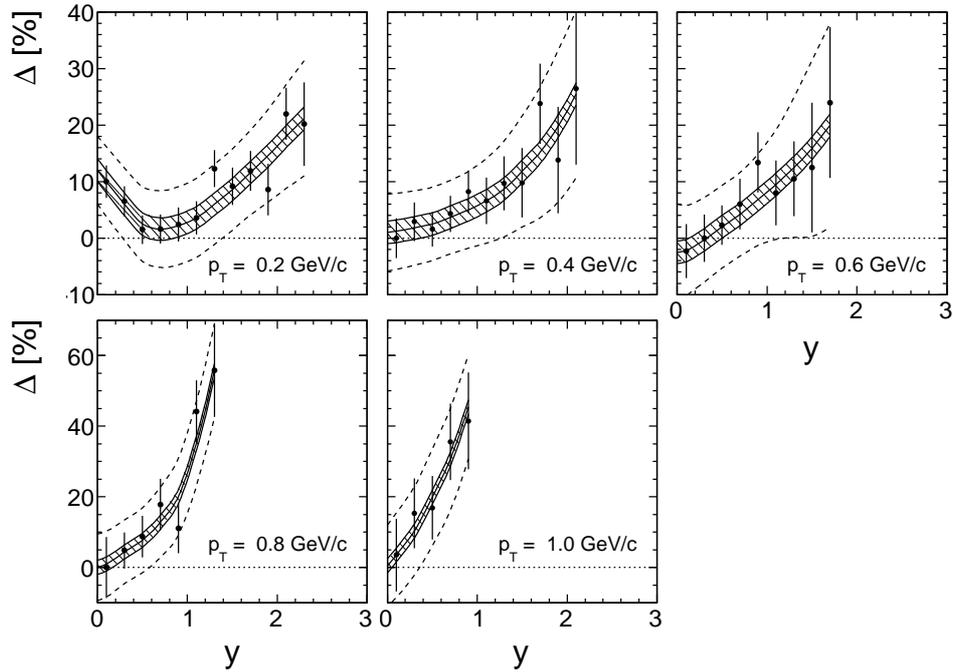}
      \caption{Relative differences between NA61 and the bubble
               chamber results \cite{blobel} interpolated to the beam momentum of
               20~GeV/c as a function of rapidity for five values of $p_T$. The full lines 
               are drawn to guide the eye. The shaded 
               areas and the dashed lines represent the systematic errors of the reference
               data and NA61 data, respectively}
  	  \label{fig:na61_bc_comp}
  \end{center}
\end{figure}

Double differential cross sections at $p_{\textrm{beam}}$~=~12 and 24~GeV/c have
been published by \cite{blobel} as functions of $p_T$ and rapidity. These
data have been interpolated to the beam momentum of 20~GeV/c
and compared to the NA61 results. The resulting relative differences
in percent are shown in Fig.~\ref{fig:na61_bc_comp}.

%        Fig.10 

Also at this beam momentum large discrepancies up to a 40--50\% level
are evident, with a systematic rapidity dependence that changes
strongly with transverse momentum.  

%
% ****************************** Section 7.3 ****************************
%
\subsection{Inverse slopes of the $m_T$ distributions ("temperature")} 
\vspace{3mm}
\label{sec:inv_slope}

\begin{figure}[h!]
  \begin{center}
  	\includegraphics[width=8cm]{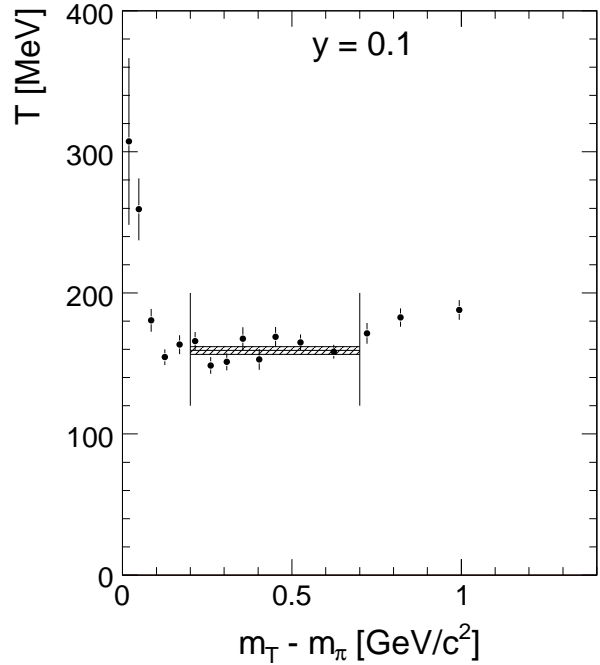}
      \caption{Local inverse slopes of the $(m_T-m_\pi)$ distribution at
               $y$~=~0.1, obtained by an exponential fit to three successive data points, 
               as a function of $(m_T-m_\pi)$ at $p_{\textrm{beam}}$~=~158~GeV/c. 
               The limits of the exponential fit are indicated by the vertical lines, the resulting
               "hadronic temperature" as the horizontal band corresponding to the published
               systematic error of $\pm$2.7~MeV}
  	  \label{fig:mt_slope}
  \end{center}
\end{figure}

NA61 has chosen to fit their $(m_T-m_\pi)$ distributions by an exponential 
within the limits 0.2~$< m_T-m_\pi <$~0.7. The lower limit is positioned well
above the mean $p_T$ such that, at central rapidity and $p_{\textrm{beam}}$~=~158~GeV/c, 
only 44\% of all $\pi^-$ fall within this $(m_T-m_\pi)$ bin. The inverse
slope of a supposedly exponential $m_T$ distribution is generally
connected to a thermodynamic quantity called "hadronic temperature".
It may be asked what relation an exponential fit over this small
$m_T$ interval may have to this quantity. Indeed, the dependence of
the inverse slope on $m_T$ may be extracted from the published NA61
data by using a local exponential fit to three successive data points. This is shown 
in Fig.~\ref{fig:mt_slope} for the most central rapidity bin at $p_{\textrm{beam}}$~=~158~GeV/c.

%        Fig.11 

\begin{figure}[h!]
  \begin{center}
  	\includegraphics[width=8cm]{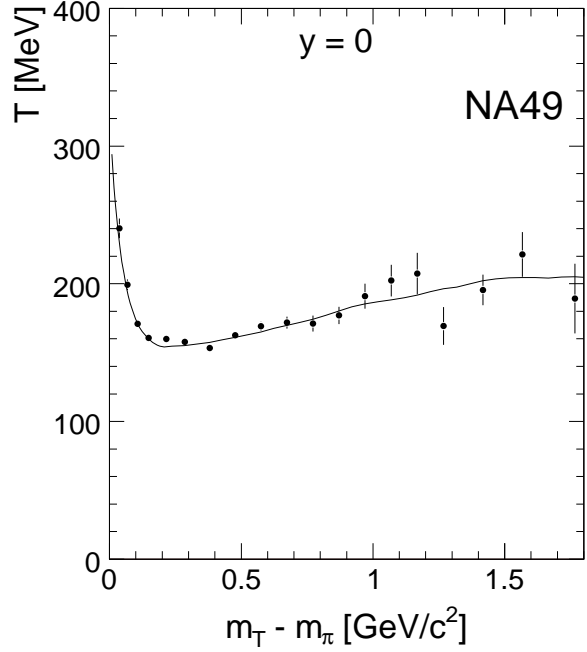}
      \caption{Inverse slopes of the $(m_T-m_\pi)$ distribution at
               $y$~=~0 as a function of $(m_T-m_\pi)$ for the NA49 data \cite{pp_pion}}
  	  \label{fig:inv_slope_na49}
  \end{center}
\end{figure}

A look at Fig.~\ref{fig:mt_slope} shows that the fit limits appear to have been chosen
to cover mainly the region of the $m_T$ dependence that might be called
exponential within the sizeable statistical bin-by-bin errors.
Immediately below and above these limits the inverse slopes increase
towards low and high $m_T$. In fact the $(m_T-m)$ distributions
have been shown to be non-exponential for pions \cite{pp_pion}, kaons \cite{pp_kaon} and
baryons \cite{pp_proton} by the NA49 collaboration. This is very evident looking
at the corresponding inverse slopes for $\pi^-$ as a function of $(m_T-m_\pi)$
as published, with an event number three times higher than NA61,
by NA49 \cite{pp_pion}, see Fig.~\ref{fig:inv_slope_na49}.

%        Fig.12 

\begin{figure}[h!]
  \begin{center}
  	\includegraphics[width=8cm]{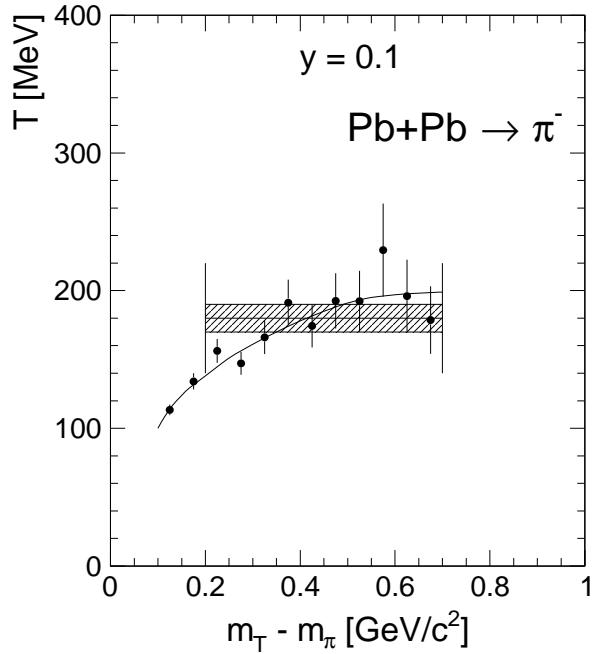}
      \caption{Local inverse slopes of the $m_T - m_\pi$ distribution
               at $y$~=~0.1 for central Pb+Pb interactions \cite{pbpb} at
               $p_{\textrm{beam}}$~=~158~GeV/c. The full line is drawn to
               guide the eye, the horizontal line gives the claimed temperature
               in the limits 0.2~$< m_T - m_\pi <$~0.7~GeV/c$^2$ with the shaded
               area corresponding to the systematic error}
  	  \label{fig:mt_slope_pb}
  \end{center}
\end{figure}

Indeed there is practically no region in this plot where the inverse slopes
may be called constant. Between the cut limits 0.2~$< (m_T-m_\pi) <$~0.7
of the NA61 analysis there is a systematic change of the inverse
slopes by about 15~MeV, to be compared to the given systematic
error of the NA61 fit of only 2.7~MeV. In this context also the comparison of
the inverse slopes between central Pb+Pb and p+p interactions \cite{na61} (Fig.~18) 
needs comment. The published Pb+Pb data \cite{pbpb} show a very strong dependence
of the inverse slopes on $m_T - m_\pi$, see Fig.~\ref{fig:mt_slope_pb}.

%        Fig.13 

In view of a variation of the inverse slopes of about 100~MeV in the interval 
0.1~$< m_T - m_\pi <$~0.7~GeV/c$^2$ the very concept of a well-defined and
constant "hadronic temperature" must be questioned.

The value of 180~MeV quoted in \cite{na61} (Fig.~18) with a systematic error
of only 10~MeV, will change to 163~MeV by changing the fit interval from 
0.2~$< m_T - m_\pi <$~0.7~GeV/c$^2$ to 0.1~$< m_T - m_\pi <$~0.6~GeV/c$^2$
while the same change in the p+p data will keep the $T$ value stable to within
1~MeV. In addition it should be recalled that the inverse slopes have to
diverge upwards in the approach to $m_T - m_\pi$~=~$p_T$~=~0 as the
invariant cross section crosses this value with slope zero.

%
% ****************************** Section 8 ****************************
%
\section{Corrections and normalization} 
\vspace{3mm}
\label{sec:corr}

Given the large discrepancies between the NA61 results and the
set of reference experiments described above, it should be 
mandatory to localize apparative and experimental effects that
might be responsible for the observed problems. This would
concern in particular the size of the corrections to the
raw data and the way the results are normalized. Unfortunately
there are no quantitative informations available in \cite{na61}
with respect to these important ingredients to the data analysis.
The reader might use the repeated statement that the systematic
errors are enumerated as 20-40\% of the applied corrections to
estimate that the corresponding corrections might reach values
of up to or bigger than 50\%. Concerning the absolute normalization
the statement concerning a change of the target density with
time in Sect.~3 of the paper \cite{na61} would indicate that the standard
normalization via a model-independent direct measurement of
the trigger cross section \cite{pp_pion} was impossible, see also the
comments in \cite{na61_prog}. Instead the normalization has to
rely completely on the comparison to a microscopic model \cite{epos}.
The event losses due to the interaction trigger of NA61 are
almost completely located in the diffractive sector of the
strong interaction, a sector that is notoriously ill described
by the standard microscopic models as those do not contain the
production and cascading decays of N$^*$ resonances which dominate
this region of phase space. Furthermore, the used production model will
not only have to describe exactly the inclusive yields of forward
particles hitting the trigger counter, but also the correlation of
these particles with secondary $\pi^-$ in order to quantify the correction
to be applied at lower rapidity. This correction is strongly dependent on
$p_T$ and $y$ \cite{pp_pion}. In this context the statement that
"the results presented in this paper are determined from particle
yields per selected event" does not take into account that all reference
yields discussed in the present comments are referred to the
total inelastic cross section and not to an undetermined trigger
cross section.

%
% ****************************** Section 9 ****************************
%
\section{Conclusion} 
\vspace{3mm}
\label{sec:conclusion}

New data on negative pion production from the NA61 collaboration
in the range of beam momenta from 20 to 158~GeV/c have been compared
to a wide range of reference data in the same energy region. 
Important deviations of the new data from the existing results
have been revealed. These discrepancies increase in the
successive steps from the total pion yields via the single
differential rapidity distributions to the double differential
invariant cross sections.

\end{document}